\documentclass[aip,rsi,amsmath,amssymb,longbibliography,superscriptaddress,reprint,nobibnotes]{revtex4-1}

\usepackage{graphicx}% Include figure files
\usepackage{dcolumn}% Align table columns on decimal point
\usepackage{bm}% bold math
\usepackage[utf8]{inputenc}
\usepackage[T1]{fontenc}
\usepackage{mathptmx}
\usepackage{etoolbox}
\usepackage{upgreek}

\usepackage{import}
\usepackage{xifthen}
\usepackage{transparent}
\usepackage{subfig}
\usepackage{xcolor}

\usepackage{siunitx}

\newcommand{\UCBChem}{Department of Chemistry, University of California, Berkeley, California 94720, USA}
\newcommand{\LBNL}{Materials Sciences Division, Lawrence Berkeley National Laboratory, Berkeley, California 94720, USA}
\newcommand{\FSU}{Institute for Optics and Quantum Electronic, Friedrich Schiller University Jena, 07743 Jena, Germany}

\makeatletter
\def\@email#1#2{%
 \endgroup
 \patchcmd{\titleblock@produce}
  {\frontmatter@RRAPformat}
  {\frontmatter@RRAPformat{\produce@RRAP{*#1\href{mailto:#2}{#2}}}\frontmatter@RRAPformat}
  {}{}
}%
\makeatother

\begin{document}

\title{A solid-state high harmonic generation spectrometer with cryogenic cooling}

\author{Finn Kohrell$^\dagger$}
\affiliation{\FSU}
\affiliation{\UCBChem}
\author{Bailey R.~Nebgen$^\dagger$}
\affiliation{\UCBChem}
\affiliation{\LBNL}
\author{Jacob A.~Spies$^\dagger$}
\affiliation{\UCBChem}
\affiliation{\LBNL}
\author{Richard Hollinger}
\affiliation{\FSU}
\affiliation{\UCBChem}
\author{Alfred Zong}
\affiliation{\UCBChem}
\affiliation{\LBNL}
\author{Can Uzundal}
\affiliation{\UCBChem}
\author{Christian Spielmann}
\affiliation{\FSU}
\author{Michael Zuerch}
\affiliation{\UCBChem}
\affiliation{\LBNL}
\email[Correspondence to: ]{mwz@berkeley.edu.          $^\dagger$These authors contributed equally.}

\date{\today}

\begin{abstract}
Solid-state high harmonic generation spectroscopy (sHHG) is a promising technique for studying electronic structure, symmetry, and dynamics in condensed matter systems. Here, we report on the implementation of an advanced sHHG spectrometer based on a vacuum chamber and closed-cycle helium cryostat. Using an in situ temperature probe, it is demonstrated that the sample interaction region retains cryogenic temperature during the application of high-intensity femtosecond laser pulses that generate high harmonics. The presented implementation opens the door for temperature-dependent sHHG measurements down to few Kelvin, which makes sHHG spectroscopy a new tool for studying phases of matter that emerge at low temperatures, which is particularly interesting for highly correlated materials.
\end{abstract}

\maketitle

\section{\label{sec:Introduction}Introduction}
\label{fund:sHHG}
\begin{figure}
    \def\svgwidth{0.95\columnwidth}
    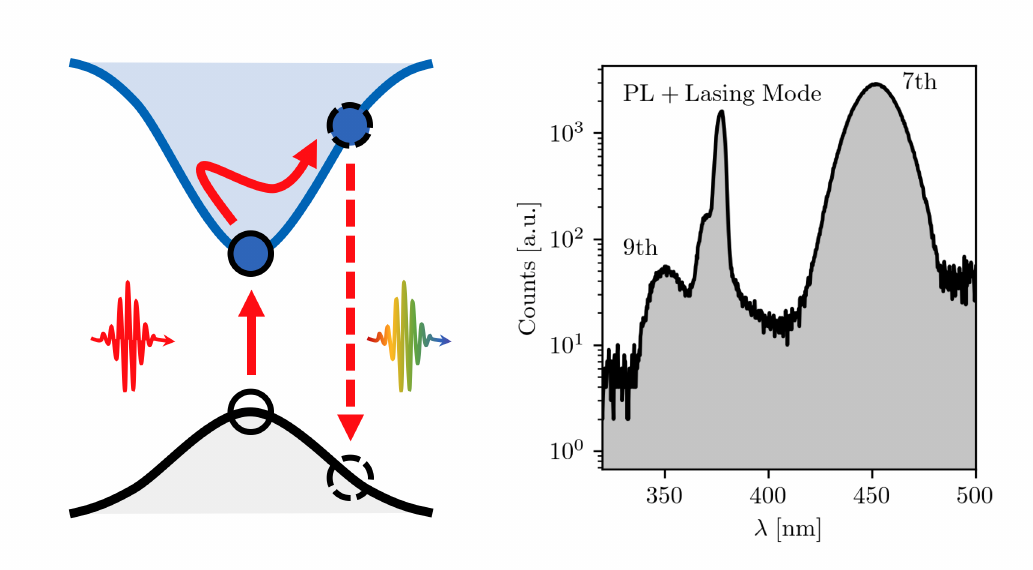

    \caption{sHHG mechanism and example ZnO spectrum. \newline \textbf{a)} Sketch showing the principle of the sHHG mechanism. \textbf{(1)} In the presence of a strong laser field, an electron is tunnel-ionized from the valence band (VB) into the conduction band (CB), creating a hole in the VB. \textbf{(2)} The electron-hole pair is accelerated along the materials nonlinear band structure, giving rise to the \textit{intraband} component of the sHHG process. \textbf{(3)} Polarization buildup between VB and CB and subsequent recombination of electron-hole pairs cause the \textit{interband} component of sHHG emission. \newline \textbf{b)} Example ZnO spectrum showing the 7th and 9th harmonic as well as broad photoluminescence (PL) peak and emerging lasing mode, taken at an intensity of $0.41 \pm 0.02~$TW/cm$^2$ and $T = 19$~K temperature.}
    \label{fig:introduction}
\end{figure}

High harmonic generation (HHG) is a highly nonlinear optical process that has become a standard method for generating ultrashort extreme ultraviolet (XUV) pulses in noble gases. In the gas phase, the process can be classically described by the three-step model, in which electrons tunnel-ionize from an atom in a strong field, accelerate through space, and then re-collide with the parent ion with excess kinetic energy, which results in harmonic emission. In the solid phase, HHG was first observed in bulk crystals in 2011 using strong mid-infrared (MIR) laser pulses to generate a visible to ultraviolet harmonic spectrum.\cite{ghimire_observation_2011} Solid-state HHG (sHHG) occurs by a similar process to gas HHG and exhibits similar features such as a plateau of harmonics reaching out to a cut-off at high energies. However, one key aspect of how sHHG deviates from the basic three-step model is that accelerated electrons are not free particles, but rather constrained by allowed bound energy states in a periodic crystal structure (see Fig.~\ref{fig:introduction}a). This allows harmonics to both be emitted from electron-hole recombination (interband polarization), as well as directly from a nonlinear intraband current of electrons in the conduction band and holes in the valence band due to Bloch oscillations.\cite{teichmann_high_2011, ghimire_observation_2011, ghimire_high-harmonic_2019} The process of accelerating charge carriers through the material's band structure accounts for major differences observed between sHHG and gas HHG spectra and makes sHHG a promising tool for studying the electronic structure of solids, as the potential energy surface experienced by accelerated electrons is encoded in the efficiency of harmonic emission in the sHHG spectrum. The potential energy surface and therefore the harmonic spectrum also depends on the lattice direction along which electrons and holes are accelerated, providing a way to access 3D band structure through measuring the anisotropy of harmonics, which has become a standard measurement in the field. sHHG has been observed in a broad variety of materials, from wide-gap dielectrics like SiO$_2$ \cite{luu_extreme_2015} and GaSe \cite{ndabashimiye_solid-state_2016} to \mbox{(quasi-)2D} materials like graphene\cite{Bowlan2014, Taucer2017, Yoshikawa2017, Baudisch2018} and transition metal dichalcogenides\cite{Liu2017, Lou2020, Yoshikawa2019, Heide2022, Yue2022c} and has inherent capabilities to probe phenomena on femtosecond timescales and beyond,\cite{Bowlan2014, Langer2016, Garg2016, Uzan2020, Bionta2021, Heide2022} making it an emerging probe with significant potential to study electronic structure and properties in solids.\cite{Zong2023}

sHHG has recently been extended to investigations of new and exotic classes of quantum materials such as Weyl semimetals\cite{Lv2021} and topological insulators.\cite{Baykusheva2021, Schmid2021, Heide2022b} These strongly correlated systems often exhibit unique phase transitions that can be accessed through changes in external conditions, in particular, a large number of relevant materials require access to cryogenic temperature to reach such phase transitions. To this end, we have implemented a sHHG spectrometer with a vacuum chamber and closed-cycle helium cryostat with the goal of measuring temperature-dependent sHHG spectra at temperatures down to $\sim10$~K and pressures down to $10^{-10}$~Torr, allowing the study of electronic band structure in hard-to-access phases and across phase transitions. In particular, to precisely analyze temperature-dependent dynamics of thin film materials using sHHG spectroscopy, accurate determination of the sample temperature in the optical interaction region is crucial. Given the high intensities required to drive the sHHG process and mW level average powers applied to thin-film samples, it is critical to assess the sample temperature in situ and determine the relationship between local sample temperature relative to external temperature probes such as typically installed on sample holders. Ensuring accurate sample temperatures is crucial for avoiding experimental artifacts and potentially erroneous conclusions.\cite{Kivelson2020} As a test case to benchmark the performance of the apparatus, we investigate the band gap shift of a ZnO thin film while simultaneously measuring the 7th and 9th harmonic, as shown in Fig.~\ref{fig:introduction}b. It is found that by applying laser intensities in the Mott transition regime, a strong photoluminescence (PL) signal featuring a lasing transition, as has been previously observed in ZnO,\cite{wille_carrier_2016} one can accurately track the temperature-dependent band gap by measuring the PL wavelength and therefore utilize this emission for accurate in situ sample temperature assessment under strong-field conditions.

%%% Technical Description %%%
\section{\label{sec:Setup}Technical Description}
\subsection{Overview of Instrument}

\begin{figure*}[ht]
    \centering
    \def\svgwidth{0.95\linewidth}
    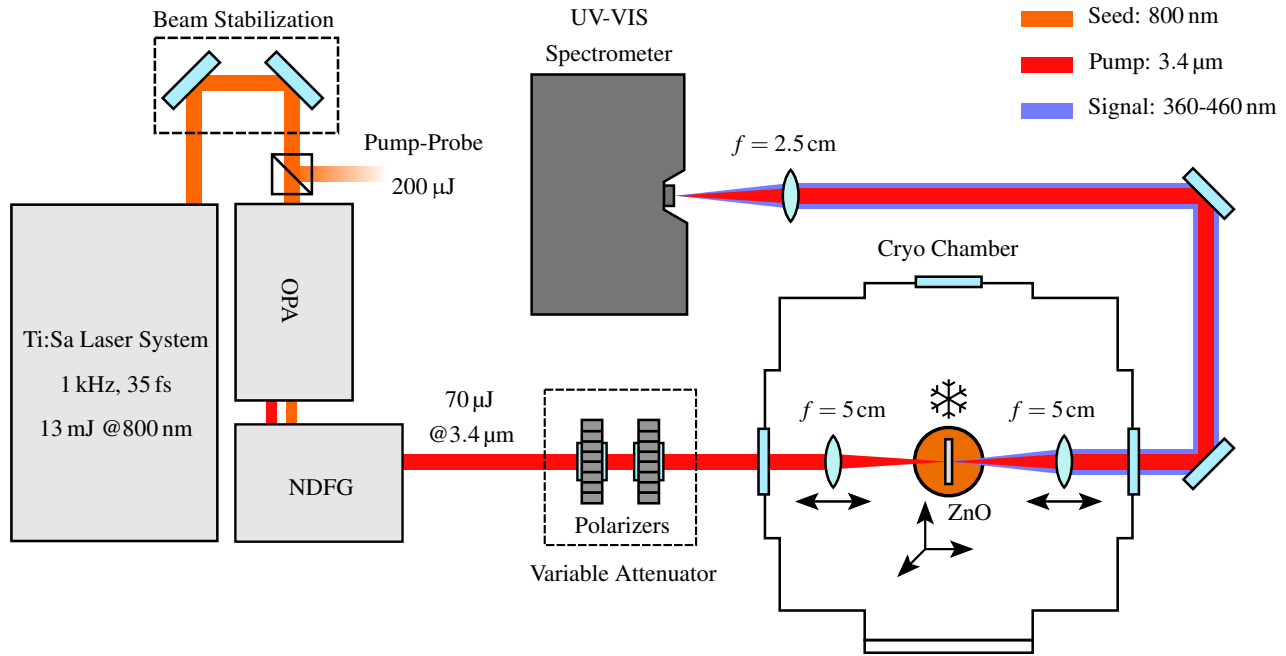

    \caption{Schematic experimental setup for the ZnO measurements. The Ti:sapphire laser system delivers 13~mJ pulses at 800~nm wavelength with a duration of 35~fs at a repetition rate of 1~kHz. 5~mJ is used to pump the optical parametric amplifier (OPA). After a beam stabilization system the pulses are sent to the OPA and noncollinear difference frequency generation (NDFG) unit, which are tuned to turn the 800-nm seed into a 3.4~$\upmu$m output with a pulse energy of about 70~$\upmu$J. A pair of wire-grid polarizers allows continuous adjustment of the transmitted pulse energy before the beam transmits into the vacuum chamber (see Fig.~\ref{fig:chamber-open}a. Additional polarization optics used for anisotropy measurements were omitted for clarity. A first lens on a single-axis translational stage (along the laser axis) focuses the beam onto the ZnO thin film, which is fixed to the sample holder on a triple-axis translational stage (see Fig.~\ref{fig:chamber-open}b). An identical lens after the sample collimates the driver and signal beams, which are subsequently transmitted out of the vacuum chamber and to the UV-VIS spectrometer for measurement.}
    \label{sketch:setup-berkeley}
\end{figure*}

A schematic diagram of the cryogenic sHHG spectrometer is shown in Fig.~\ref{sketch:setup-berkeley}. The instrument is based on a commercial Ti:sapphire chirped pulse amplifier system (Coherent Legend Elite Duo, 800~nm center wavelength, 1~kHz repetition rate, $\sim13$~mJ amplified pulse energy, $\sim35$~fs pulse duration). A portion of the amplifier (5~mJ) used for this spectrometer pumps a commercial optical parametric amplifier (Light Conversion TOPAS) and noncollinear difference frequency generation (DFG) unit (Light Conversion NDFG). An active laser beam stabilization system (MRC Systems GmbH) is used between the amplifier and optical parametric amplifier to improve shot-to-shot stability and prevent long-time alignment drift. The MIR wavelength ranges from 2.6--11~$\upmu$m with typical measurements utilizing 3--5~$\upmu$m radiation with $\sim30~\upmu$J pulse energy.

The MIR driving beam is then transmitted through a series of polarization optics consisting of a pair of wire-grid polarizers (Thorlabs WP25M-IRA) to control the field strength and an achromatic half waveplate (B.Halle Nachfl. GmbH, Achromatic 3--6~$\upmu$m) mounted in a motorized rotation stage (Thorlabs K10CR1) to control the polarization state for sHHG anisotropy measurements. The MIR driving beam is transmitted into and out of the vacuum chamber through 3~mm thick CaF$_2$ windows (Edmund Optics 11-875). The focusing and collimating optics are two uncoated plano-convex CaF$_2$ lenses with 50~mm focal lengths (Thorlabs LA5763). Due to the size of the vacuum chamber, the focusing and collimating optics are both placed inside the chamber. Both lenses are mounted on motorized linear translation stages driven by piezoelectric linear actuators (Newport PicoMotor) and their longitudinal positions are remotely controlled from the outside of the chamber, allowing them to be optimized even under vacuum conditions. This allows for a high degree of dynamic control over the effective beam size on the sample surface down to a 50~$\upmu$m diameter at the focus.

The collimated high-harmonic emission in the UV-visible range is analyzed using another wire-grid polarizer (Thorlabs WP25M-UB) mounted in a motorized rotation stage (Thorlabs K10CR) and is focused into a high-sensitivity imaging spectrometer (Andor Kymera 328i) with a Peltier-cooled camera (Andor iDus DU420A-BU2) using a 25.4~mm focal length uncoated biconvex CaF$_2$ lens (Thorlabs LB5774). A typical polarization anisotropy measurement involves measuring the resulting high-harmonic spectra both parallel and perpendicular to the MIR driving field polarization; both the MIR driving field polarization and high-harmonic emission analyzer polarizer are rotated together. However, the benchmark measurements presented in Sec.~\ref{sec:results} focus on characterizing the cryogenic capabilities of the instrument, so polarization anisotropy measurements were not performed in this work.

\subsection{\label{subsec:components}Description of the Cryostat}
\subsubsection{\label{sec:level3} Vacuum Chamber}
\begin{figure}[ht]
    \centering
    \def\svgwidth{0.95\columnwidth}
    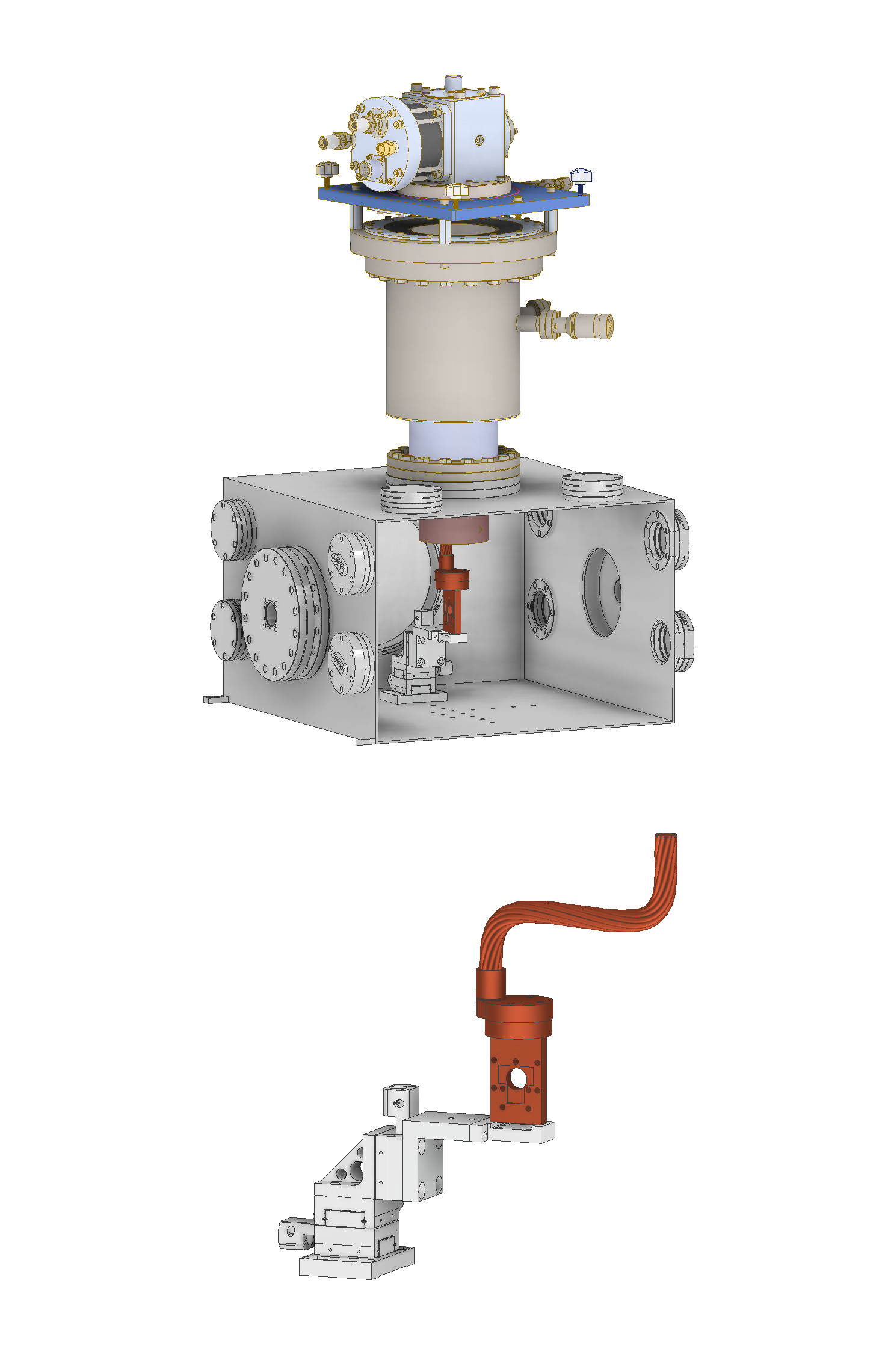

    \caption{Labeled CAD models of the vacuum chamber with closed-cycle helium cryostat and sample environment. \newline \textbf{a)} View of the complete chamber assembly including closed-cycle helium cryostat, window flanges for beam/signal transmission, and electrical feedthroughs. The optical axis (OA) is represented by a dashed red line. A quick-access loading door is mounted on the front chamber wall (removed from sketch to allow inside view). \textbf{b)} Sample holder mounted on a PEEK adapter connected to the 3D translation stage controlled by piezo linear actuators. The adapter and 1~cm diameter copper braid ensure thermal exchange with the cryostat while allowing adjustments of the sample position using the triple-axis translational stage.}
    \label{fig:chamber-open}
\end{figure}

A diagram of the custom-made vacuum chamber (Kurt Lesker) is shown in Fig.~\ref{fig:chamber-open}a. The vacuum chamber features a quick-access loading door (Kurt Lesker DS-LL0800) mounted on a DN150CF port, two DN100CF ports along the optical axis for transmission measurements, an additional DN150CF port perpendicular to the sample for reflection measurements, and ten smaller DN63CF ports for accessories such as electrical feedthroughs. Currently, the vacuum chamber is equipped with electrical feedthroughs for the piezo linear actuators, a cold cathode pressure gauge (Pfeiffer Vacuum IKR 270), and a burst disk (Ideal Vacuum Products P107370) for nitrogen venting. Ultra-high vacuum is achieved using a turbo pump (Pfeiffer Vacuum HiPace 80) backed by an oil-free dry scroll roughing pump (Edwards nxDS10i). The foreline pressure is measured using a Pirani gauge (Pfeiffer Vacuum TPR 280). A subsequent leak test showed the chamber pressure stabilizing in the low $10^{-8}$~Torr regime. By baking out the chamber, the pressure was successfully reduced by another order of magnitude. The final chamber pressure reached $\sim1.2\times10^{-9}$~Torr, which is sufficient to safely operate the liquid helium cryostat without noticeable condensation on the sample.

\subsubsection{\label{subsec:Cryo} Closed-Cycle Cryostat}
The closed-cycle helium cryostat (Sumitomo Heavy Industries, Ltd. RDK-101D AK Cold Head) depicted in Fig.~\ref{fig:chamber-open}a consists of two cooling stages that provide closed-cycle refrigeration down to 4~K. This technique relies on expansion of high-pressure helium gas at 250~psi, supplied by a helium compressor (Sumitomo Heavy Industries, Ltd. HC-4E1). This periodic compression introduces significant vibrations that can affect laser stability and precise focusing onto the sample that is attached to the cooling finger. The vibration is especially problematic for studying small samples, such as exfoliated 2D materials, where the sample size is on the order of the MIR spot size. To remedy this issue, the cooling head and cooling finger are decoupled from each other for operation and thermal exchange between them is achieved using a helium gas filled low-vibration interface. Therefore, this part of the cryostat needs an additional external helium supply for continued effective thermal exchange. The vibrational isolation interface slightly reduces the cooling power of the cryostat, and in combination with the sample's connection to the translational stage adding cooling load, the effective minimum temperature is $\sim16$~K. However, this could be remedied with additional radiation shielding and/or adjustment of the vibrational isolation interface.

\subsubsection{Sample Environment}
The sample holder is made from oxygen-free high thermal conductivity (OFHC) copper and is mounted on a three-axis linear translation stage via an adapter made out of polyetheretherketone (PEEK), a thermally insulating plastic, allowing for 3D spatial positioning of the sample during operation using a series of piezoelectric linear actuators (Newport PicoMotor). A schematic diagram of the sample holder and 3D positioning stage is shown in Fig.~\ref{fig:chamber-open} \textbf{b)}. The sample holder and the cooling finger are connected via a 11.6~mm diameter OFHC copper braid which has been crimped onto machined connector pieces that effectively connect the cooling finger tip to the sample holder through the braid. All contact surfaces were polished.

\section{\label{sec:results}Results and Discussion}

\subsection{Choice of Calibration Target}
In order to benchmark the performance of the cryogenic sHHG spectrometer described above, temperature-resolved measurements were performed on ZnO. ZnO is a direct band gap semiconductor ($E_g = 3.40$~eV) that has been proposed as a promising candidate to compete with common semiconductors such as Si, GaAs, and GaN for a broad range of applications.\cite{look_recent_2001} ZnO-based devices are significantly less expensive to fabricate than dry-etched GaN-based devices due to the well-developed bulk and epitaxial growth techniques based on wet chemical processes.\cite{pearton_recent_2003, ozgur_comprehensive_2005} They also have high breakdown strength and saturation velocity, making them appealing for electronic applications such as piezoelectric transducers and transparent conducting films.\cite{golovanov_zno_2022, zhao_investigation_2022}

However, its primary advantage, especially as a photon emitter, lies in its large exciton binding energy of about 60~meV (compared to GaN with 26~meV), which makes ZnO a highly efficient excitonic emitter even at low excitation energies and high temperatures. For over a decade, this has drawn significant attention to its potential relevance in short-wavelength optoelectronic devices such as semiconductor micro- and nanolasers, which are of interest for the construction of compact integrated circuits that are able to enhance the spatial resolution of a variety of sensors and imaging applications.\cite{rai_elevated_2012, dai_electronhole_2014, wille_carrier_2016, sentosa_temperature_2011}

For prospective use as a semiconductor laser material, the density of free carriers is crucial because it effectively determines the optical properties. Spontaneous emission leading to photoluminescence due to electron-hole recombination in excitons takes place at fluences below the threshold excitation density, for which the carrier density is low and the radiative decay times are long ($\sim150$~ps). Exceeding this critical threshold, when the carrier density reaches the level characterized by the Mott density, the excitons scatter and eventually dissociate into unbound electrons and holes due to screening of their binding potential. They form an electron-hole plasma accompanied by much higher scattering rates and thus faster radiative decay ($\sim5$~ps), which causes the semiconductor to exhibit strong stimulated emission and therefore enables lasing at $\sim400$~nm through the recombination of the free electrons and holes.\cite{dai_electronhole_2014, wille_carrier_2016} Because of its high sensitivity to the charge carrier density in the conduction bands, sHHG spectroscopy presents a promising new approach to reveal insights into charge carrier dynamics around the Mott transition. In this study, we use a well characterized material, ZnO, to show the technique's capabilities and its potential for investigation of less well understood materials.

In addition to potential technological relevance, ZnO is an ideal spatio-temporal alignment target owing to its strong emission in the UV-visible region that can be used to align and optimize the instrument. Furthermore, the well-studied temperature-dependent band gap enables verification of cryogenic temperatures under experimental conditions (e.g., MIR irradiation). ZnO exhibits a blue-shift of the band gap upon cooling like other semiconductors, leading to an observed shift of the photoluminescence and lasing mode to shorter wavelength. This is due to electron-phonon interactions\cite{Fan1950, Muto1950, Fan1951} and a decrease in inter-atomic spacing with decreasing thermal energy, which consequently increases the average potential seen by the electrons and therefore the band gap.\cite{rai_elevated_2012} The band gap energy as a function of temperature exhibits quadratic behavior which can be modeled empirically with the Varshni equation,
\begin{equation}\label{eqn:varshni}
    E_g(T) = E_g(0~\text{K}) - \frac{\alpha T^2}{T + \beta},
\end{equation}
where $E_g(0~\text{K})$ is the band gap at $T=0$~K and fitting parameters are $\alpha$ and $\beta$, with $\beta$ being proportional to the Debye temperature.\cite{varshni_temperature_1967}

Temperature-resolved sHHG measurements were collected on a ZnO thin film (300~nm polycrystalline film grown by RF-magnetron sputtering on a 500~$\upmu$m sapphire substrate)\cite{hollinger_polarization_2019} in the temperature range of 19~K to 285~K with MIR intensities both above and below the Mott transition ($0.473 \pm 0.015$~TW/cm$^2$ and $0.360 \pm 0.012$~TW/cm$^2$, respectively). The field strength values on target were calculated by measuring the incident pulse energy directly in front of the chamber window, accounting for its transmission, assuming a 100~fs pulse duration, and assuming a 50~$\upmu$m spot size in diameter, which had been determined in previous experiments on the same beamline with the same focusing lens. Spectra were collected from 270~nm to 530~nm using a UV-blazed grating (300~nm blaze, 300~lines/mm), which captured the 7th harmonic, 9th harmonic, and PL features simultaneously. Starting from 20~K, multiple sHHG spectra were taken at each temperature point around the Mott transition for different incident MIR field strengths. Following initial measurements of the damage threshold throughout the full temperature range, care was taken to stay below the lowest determined value of about 0.6~TW/cm$^2$ for the data reported here.

\subsection{Temperature Calibration Below the Mott Transition}
\begin{figure}[ht]
    \centering
    \includegraphics[width=\columnwidth]{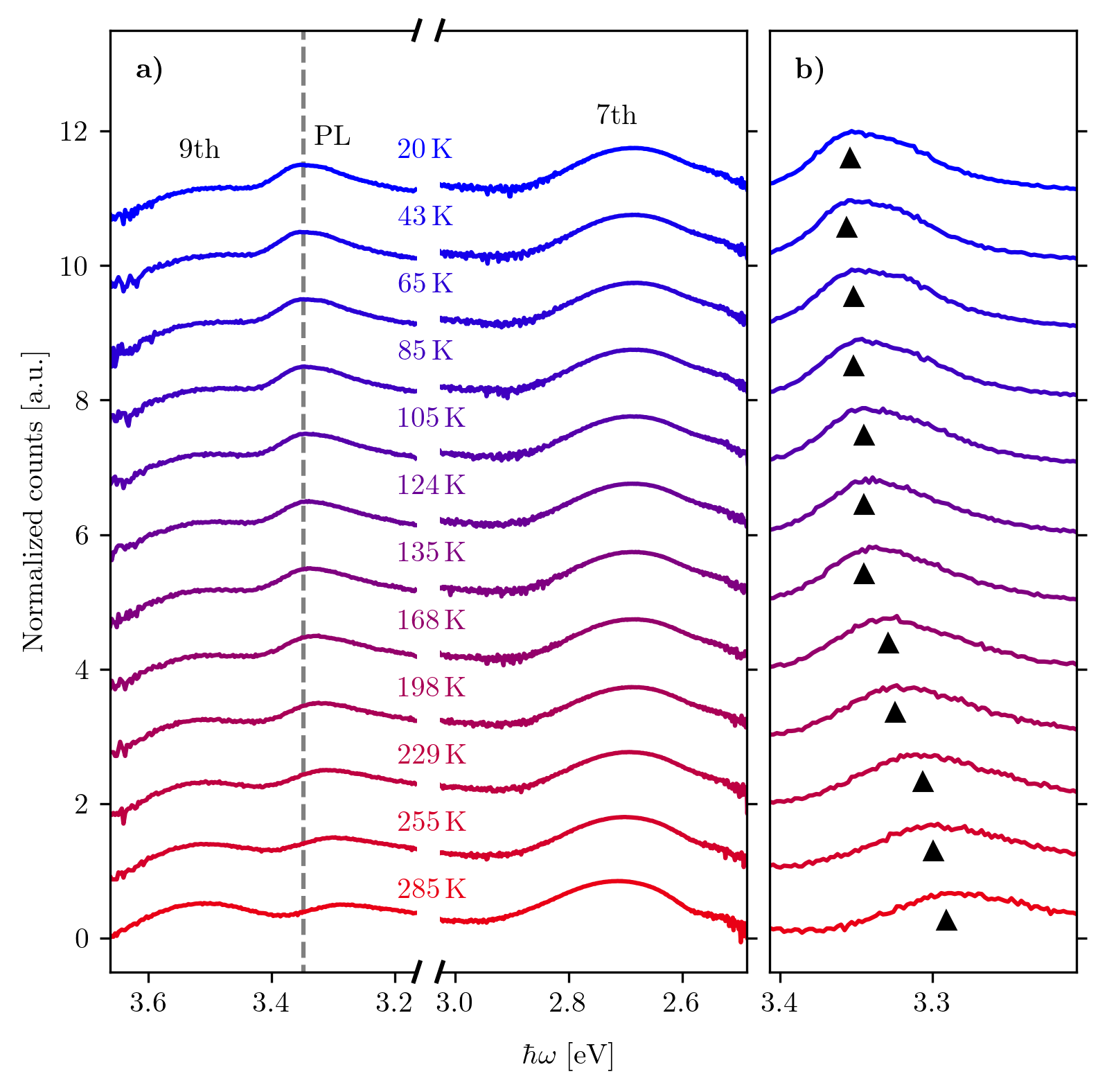}
    \caption{Normalized ZnO thin film spectra collected at different temperatures below the Mott transition at $0.360 \pm 0.012$~TW/cm$^2$ vertically offset for clarity. \textbf{a)} Spectra plotted on a logarithmic scale showing the photoluminescence (PL), 7th harmonic, and 9th harmonic signals. The dashed line indicates the peak position at the lower temperature limit (20 K) to illustrate the shift. \textbf{b)} Spectra of the PL feature used in determining the band gap of ZnO plotted on a linear scale. The center of mass values are marked to show the temperature-dependent shift of the PL.}
     \label{fig:ZnO-PL-peak-shift-below-Mott}
\end{figure}

\begin{figure}[ht]
    \centering
    \includegraphics[width=0.95\columnwidth]{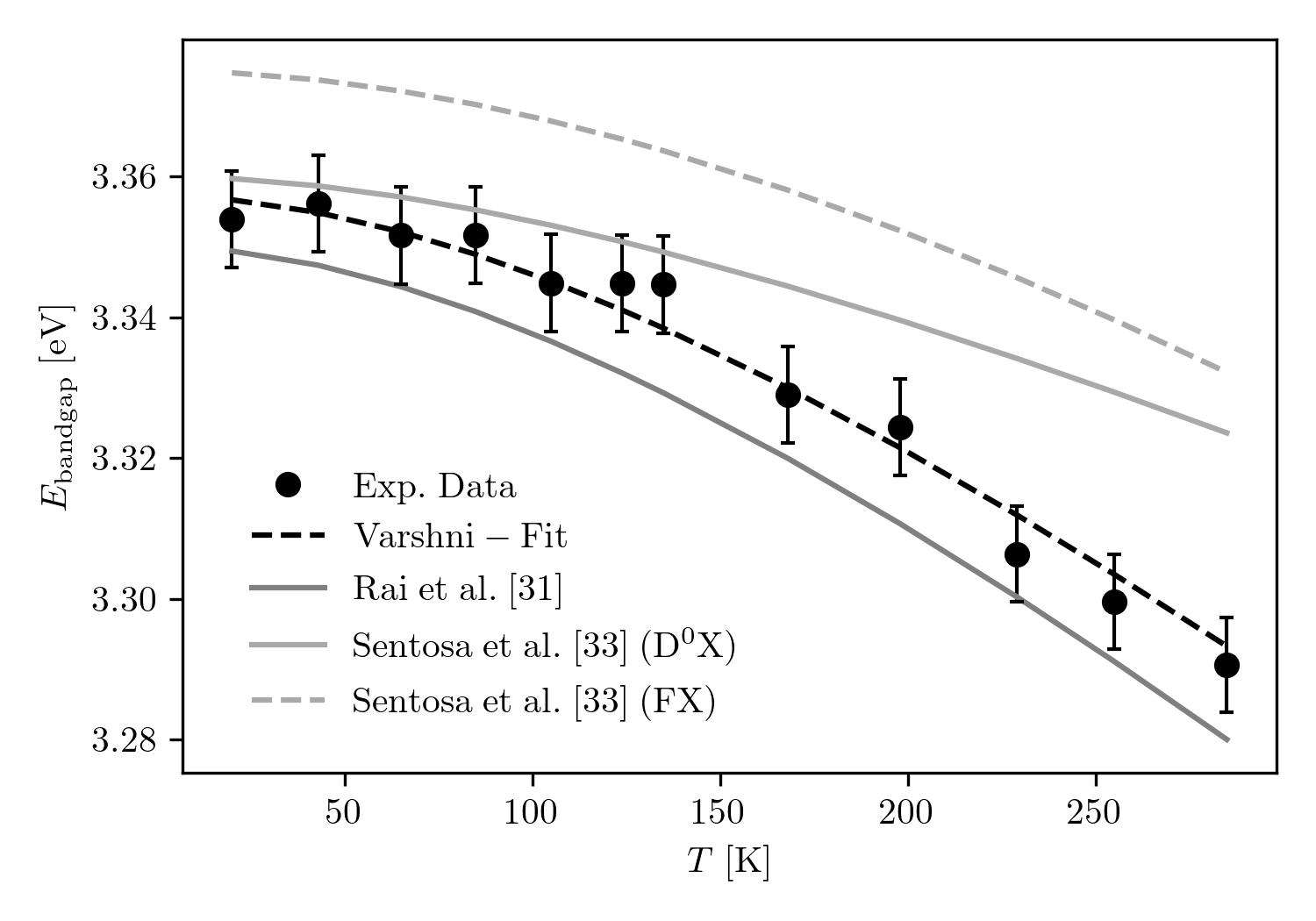}
    \caption{Measured temperature-dependent shift of the combined FX-D$^0$X feature from ZnO photoluminescence emission at $0.360 \pm 0.012$~TW/cm$^2$ laser intensity on the sample (below the Mott transition). The data was fit with the Varshni equation (Eq.~\eqref{eqn:varshni}). The other three traces are fits using the parameters determined and reported by Rai et al.\cite{rai_elevated_2012} and Sentosa et al. \cite{sentosa_temperature_2011}, where the behavior of FX and D$^0$X was investigated individually. See Table~\ref{tab:varshni_comparison} for the fit parameters.}
    \label{fig:Zno-PL+varshni}
\end{figure}

Because the photon energy of the lasing mode that appears at higher MIR intensities above the Mott transition is not necessarily the same as the band gap energy, measurements of the temperature-dependent band gap needed to be collected below the Mott transition threshold. This low-intensity experiment enables us to use the PL feature, assigned to the combined free exciton (FX) and donor-bound exciton (D$^0$X) feature, as a proxy for the band gap energy as previously utilized by Sentosa et al.\cite{sentosa_temperature_2011} Data corresponding to a MIR intensity below the Mott threshold ($0.360 \pm 0.012$~TW/cm$^2$) are shown in Fig.~\ref{fig:ZnO-PL-peak-shift-below-Mott} with peak centers marked to illustrate shifting PL. The center photon energies of the PL peaks are marked in Fig.~\ref{fig:ZnO-PL-peak-shift-below-Mott}b by solid triangles. The corresponding photon energy of the PL emission peak below the Mott transition was determined with an accuracy equivalent to $\pm 1$~nm of the center wavelength and is plotted in Fig.~\ref{fig:Zno-PL+varshni}. The data were fit to the Varshni model shown in Eq.~\eqref{eqn:varshni}, which is plotted together with three fits based on the Varshni equation from literature as a reference. Given the relative insensitivity of the fit to comparably large variations of the parameters $\alpha$ and $\beta$, they were chosen as $\alpha = 0.50$~meV/K and $\beta = 350$~K, within the error range of the values reported by Rai et al.,\cite{rai_elevated_2012} since we are measuring the same combined exciton feature. The fit using those values delivered $E_g(0~\text{K}) = 3.357 \pm 0.001$~eV, and reproduces the dependence of the measured values very well.  A comparison of the Varshni model fit parameters is shown in Table~\ref{tab:varshni_comparison}.

\begin{table}
\caption{\label{tab:varshni_comparison} Fit parameters to the Varshni formula and reported literature values for comparison. The values for $E_g$(0~K) marked with an asterisk were not reported in the corresponding paper and estimated from the respective figures.}
\begin{tabular}{||c|c|c|c||}
    \hline
     & $E_g$(0~K) [eV] & $\alpha$ [meV/K] & $\beta$ [K] \\
     \hline \hline 
     D$^0$X (Sentosa et al.\cite{sentosa_temperature_2011}) & 3.36* & 0.28 & 339\\
     FX (Sentosa et al. \cite{sentosa_temperature_2011}) & 3.375* & 0.51 & 682\\
     %Energy Gap (Rai\cite{rai_elevated_2012}) & $3.516 \pm 0.0002$ & $0.20 \pm 0.02$ & $325 \pm 20$\\
     Absorption edge (Rai et al.\cite{rai_elevated_2012}) & 3.35* & $0.53(5)$ & $330(20)$\\
     \hline
     PL emission (this work) & $3.357(1)$ & $0.50$ & $350$ \\
     \hline

\end{tabular}
\end{table}

The experimental values measured in this work shown in Fig.~\ref{fig:Zno-PL+varshni} are in reasonable agreement with results reported previously in the literature,\cite{rai_elevated_2012, sentosa_temperature_2011} especially at lower temperatures. The fit parameters reported by Rai et al. were determined from fitting data corresponding to the absorption edge of the Urbach tail for an electron-beam deposited ZnO thin film on a sapphire substrate to the Varshni equation.\cite{rai_elevated_2012} The Urbach tail lies energetically below the FX absorption peak, which is consistent with the results reported by Sentosa et al.\cite{sentosa_temperature_2011} Comparing the experimental results shown above with Sentosa et al. reveals that both of their fits are vertically shifted to higher energies. The D$^0$X trace agrees with our results at low temperatures but deviates significantly at $\sim130$~K. Above $\sim130$~K, the shape of their FX fit would better reproduce the results reported here if shifted to lower energies. This may be due to the D$^0$X contributions being dominant in the PL below 60~K, the emergence of the FX contribution at $\sim80$~K, and the subsequent vanishing of D$^0$X PL contribution at $\sim120$~K.\cite{jen_temperature-dependent_2005, sentosa_temperature_2011} Therefore, this suggests that our measurement of the combined D$^0$X and FX PL can reproduce their individual contributions in their corresponding dominant temperature regimes.

One explanation of the observed shift in the results presented in this work to lower energies may pertain to sample preparation. Sentosa et al. used a ZnO thin film sample that was prepared by pulsed laser deposition on yttria-stabilized zirconia.\cite{sentosa_temperature_2011} The thin film studied here and the one reported by Rai et al.\cite{rai_elevated_2012} were deposited on sapphire, which has a larger lattice mismatch with ZnO. The larger lattice mismatch can induce strain that has been shown to modulate the band gap in ZnO nanostructures\cite{Li2010} and promotes defects in the thin films. Furthermore, the growth technique used by Rai et al. (reactive e-beam deposition) was not identical to the one used to prepare the ZnO thin film studied here. Due to previously reported importance of growth parameters for the resulting thin film quality and subsequent electrical and optical properties,\cite{sentosa_temperature_2011} using different deposition techniques is likely to impact the observed sample behavior. Therefore the deviations observed preclude the ability to use ZnO as a quantitative spectroscopic temperature calibration. This is mainly due to the uncertainty of the values we measured for the PL feature and the relatively small shift in energy of the emission feature with temperature. This might be improved by using a ZnO sample with a well-defined structure and fabrication process such as a single crystal, allowing more precise determination of the band gap energy.

% Above Mott threshold results and discussion
\subsection{Temperature Dependence of the Mott Threshold}
\begin{figure}[ht]
    \centering
    \includegraphics[width=\columnwidth]{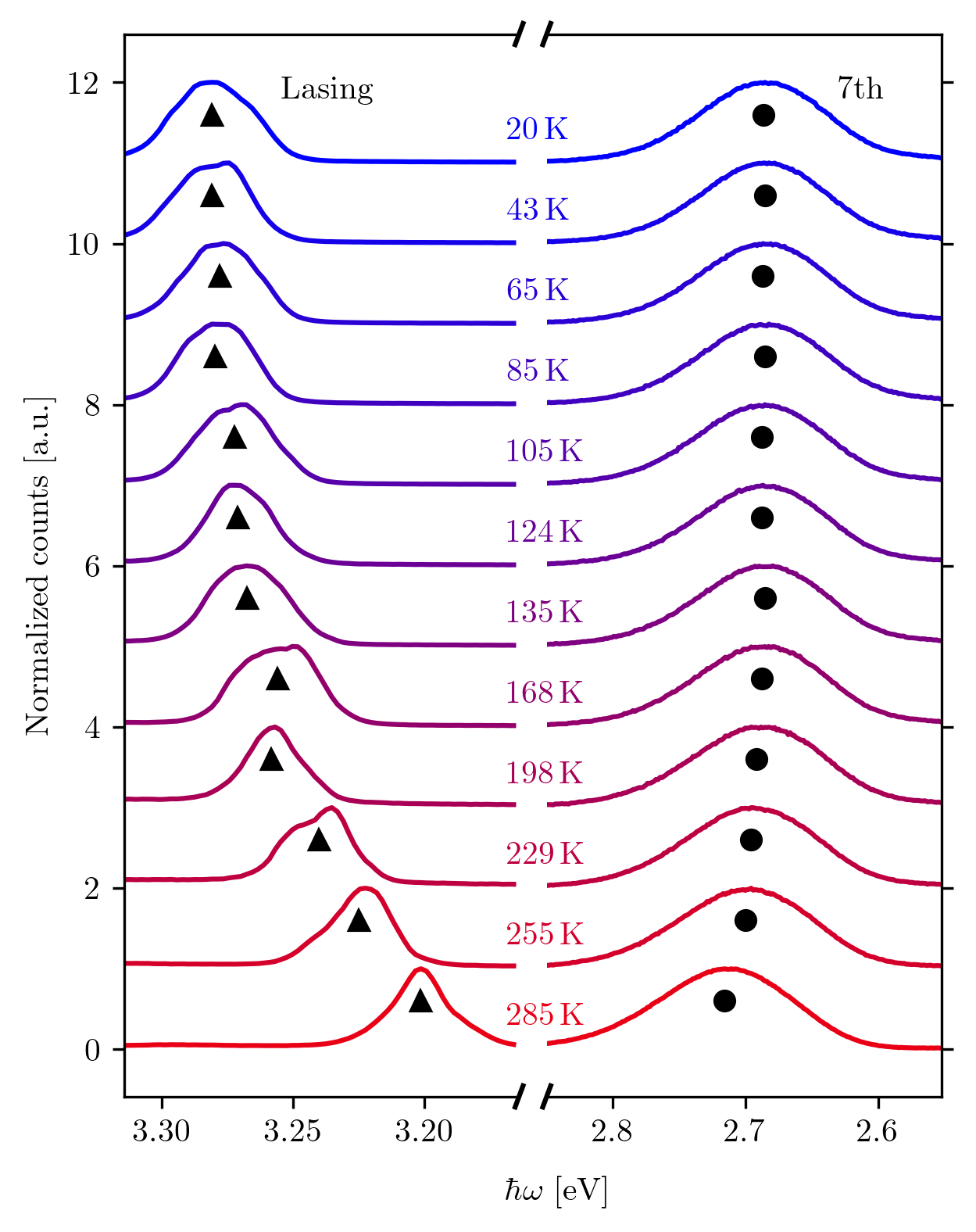}
    \caption{Normalized ZnO thin film spectra at different temperatures showing the lasing mode (left) and 7th harmonic signal (right) taken above the Mott transition at $0.473 \pm 0.015$~TW/cm$^2$. Above the Mott transition, ZnO sample exhibits a strong lasing mode in addition to harmonic emission. The spectra have been vertically offset and the center of mass values marked to show the strong temperature-dependent shift of the lasing mode and slight shift of the harmonic signal.}
    \label{fig:Zno-PL-peak-shift}
\end{figure}

In addition to the aforementioned measurements collected below the Mott transition that were utilized to calibrate the ability of the instrument to perform in the cryogenic regime, temperature- and intensity-dependent measurements were performed to directly probe the temperature-dependence of the Mott threshold, illustrating the utility of the instrument in studying important phase transitions in solids. The spectra shown in Fig.~\ref{fig:Zno-PL-peak-shift} were collected analogously to the spectra shown in Fig.~\ref{fig:ZnO-PL-peak-shift-below-Mott} at a MIR intensity above the Mott threshold ($0.473 \pm 0.015$~TW/cm$^2$). The lasing mode (solid triangles) clearly depicts an energy shift as a function of temperature, while the harmonics (solid circles) show little deviation as a function of temperature which we ascribe to slight thermal drifts of photon energy of the driver.

Measurements were also collected both as a function of MIR intensity and as a function of temperature to show how the instrument can be used to probe the Mott transition in materials such as ZnO. The intensity dependence of the 7th harmonic, 9th harmonic, and PL/lasing mode are shown below in Fig.~\ref{fig:Zno-powerlaw-293K}, illustrating the temperature dependence of the Mott transition, which can be seen as a jump in the ZnO PL intensity, corresponding to the emergence of lasing. Notably, the transition between PL and lasing occurs at a slightly lower intensity at lower temperature. Due to the PL signal below the Mott transition being an order-of-magnitude higher at 17~K, the MIR intensity could be lowered further than at room temperature without the signal dropping below noise level. Overall, these measurements illustrate the utility of sHHG spectroscopy performed under cryogenic conditions to study phenomena such as Mott transitions.

\begin{figure}[ht]
    \centering
    \includegraphics[width=0.95\columnwidth]{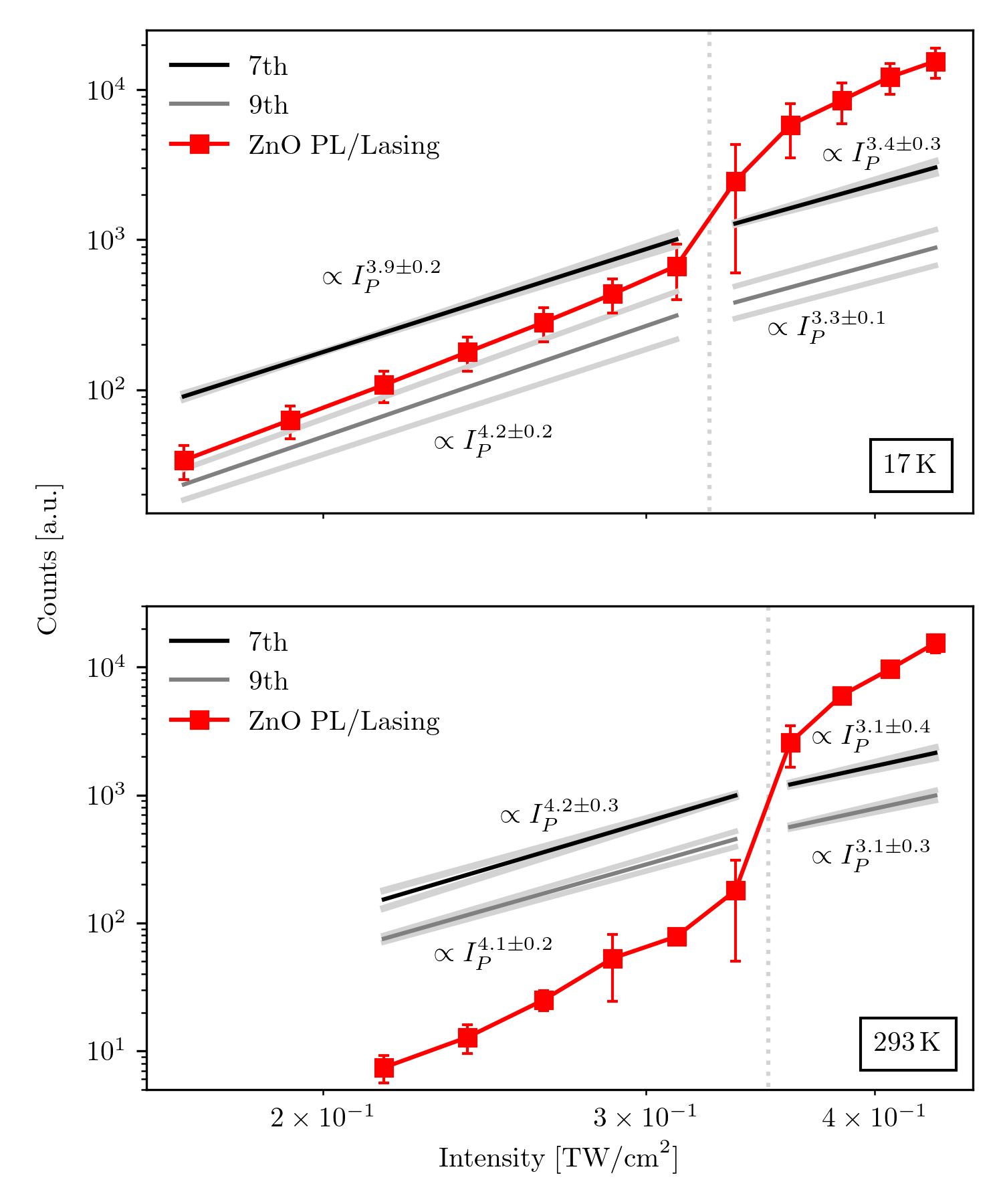}
    \caption{Intensity-dependent scaling of the 7th and 9th harmonic yield as well as the PL/lasing intensity around the Mott transition for 17~K (top) and 293~K (bottom). The individual traces and power law fits for both harmonics were averaged and are displayed including their standard deviation. The grey dashed vertical lines represent the transition from PL to the lasing regime (Mott transition), marked by a significant jump in measured peak intensity at the band gap energy.}
    \label{fig:Zno-powerlaw-293K}
\end{figure}

\section{\label{sec:conclusion}Conclusion and Outlook}
We report the development of a cryogenic sHHG spectrometer based on a closed-cycle helium cryostat. In its present form, the instrument is capable of performing measurements down to $\sim16$~K, with the possibility to reach lower temperature through radiation shielding. In order to benchmark the performance of the instrument, the temperature-dependent shift of the band gap in ZnO was investigated by extracting the peak photon energy of the PL from below threshold-density spectra and fitting the peak photon energies as a function of temperature with the Varshni formula. While there remains uncertainty due to inconsistencies regarding the temperature-dependent gap in ZnO in the available literature references, our results indicate  that it is possible to maintain cryogenic temperatures in the optical interaction region while illuminating the ZnO thin film with an intensity sufficient to reach the non-perturbative regime and generate strong sHHG signal. This indicates that sample heating under the ultrafast MIR driving field is not significant enough to preclude cryogenic measurements, resolving a debate in the community about the effective sample temperature under strong-field illumination by femtosecond pulses. This opens the door to investigate a broad range of material dynamics at low temperatures using the unique capabilities of sHHG spectroscopy.

\begin{acknowledgments}
 The authors would like to thank Alexander~Koch and Prof.~Carsten~Ronning at the Friedrich Schiller University Jena for providing the ZnO thin film used in this work. F.K. acknowledges support from the Thesis Research Fellowship Program of the German Academic Exchange Service (DAAD). B.N. acknowledges funding by the National Science Foundation Graduate Research Fellowship Program. J.A.S. acknowledges support by the Arnold O. Beckman Postdoctoral Fellowship Program. A.Z. acknowledges support from the Miller Institute for Basic Research in Science. R.H. acknowledges support by the Alexander von Humboldt Foundation. M.Z. acknowledges funding by the W. M. Keck Foundation, funding from the UC Office of the President within the Multicampus Research Programs and Initiatives (M21PL3263), and funding from Laboratory Directed Research and Development Program at Berkeley Lab (107573 and 108232).
\end{acknowledgments}

\section*{\label{sec:contribution}Author Declarations}

\subsection{Conflict of Interest}
The authors have no conflicts to disclose.

\subsection{Author contributions}
Finn Kohrell: Methodology (lead), Writing - original draft (lead), Data curation (lead), Visualization (lead), Formal Analysis (lead), Conceptualization (equal). Bailey Nebgen: Methodology (equal), Data Curation (equal), Formal Analysis (equal), Writing - review and editing (equal), Visualization (equal). Jacob Spies: Writing - review and editing (equal), Formal Analysis (equal), Visualization (equal). Richard Hollinger: Conceptualization (lead). Alfred Zong: Data Curation (equal), Conceptualization (equal), Writing - review and editing (equal), Visualization (supporting), Formal Analysis (supporting). Can Uzundal: Conceptualization (supporting), Methodology (supporting). Christian Spielmann: Supervision (supporting), Conceptualization (supporting), Writing - review and editing (supporting). Michael Zuerch: Supervision (lead), Funding acquisition (lead), Conceptualization (equal), Writing - original draft (equal), Writing - review and editing (equal).

\section*{Data Availability Statement}
The data that support the findings of this study are available from the corresponding author upon reasonable request.

\section*{references} 
%\bibliography{ref}

\end{document}